\documentclass[poms,nonblindrev]{poms1_arxiv} 

\OneAndAHalfSpacedXI

\usepackage{graphicx}
\usepackage{subfig}
\usepackage{svg}

\usepackage{natbib}
 \bibpunct[, ]{(}{)}{,}{a}{}{,}
 \def\bibfont{\small}

\usepackage{hyperref}
\usepackage{cleveref}

\usepackage{dsfont}
\newcommand{\Ic}{\ensuremath{\mathcal{I}}}
\newcommand{\Lc}{\ensuremath{\mathcal{L}}}

\newcommand{\Tc}{\ensuremath{\mathcal{T}}}
\newcommand{\Uc}{\ensuremath{\mathcal{U}}}

\newcommand{\Rd}{\ensuremath{\mathds{R}}}
\TheoremsNumberedThrough   
\ECRepeatTheorems
\EquationsNumberedThrough

\usepackage{longtable}
\usepackage{booktabs}
\usepackage{multirow}
\usepackage{array}
\newcolumntype{L}{>{$}l<{$}}
\newcolumntype{C}{>{$}c<{$}}
\newcolumntype{R}{>{$}r<{$}}

\usepackage{makecell}

\usepackage{url}
\usepackage{enumerate}
\usepackage{enumitem}
\usepackage{eurosym, relsize}
\begin{document}
\RUNAUTHOR{Qorbanian, Löhndorf, and Wozabal}

\RUNTITLE{Valuation of Power Purchase Agreements}

\TITLE{Valuation of Power Purchase Agreements for Corporate Renewable Energy Procurement}

\ARTICLEAUTHORS{%
\AUTHOR{Roozbeh Qorbanian}
\AFF{Luxembourg Centre for Logistics and Supply Chain Management, University of Luxembourg, Luxembourg, L-1359, Luxembourg, \EMAIL{roozbeh.qorbanian@uni.lu}}

\AUTHOR{Nils Löhndorf}
\AFF{Luxembourg Centre for Logistics and Supply Chain Management, University of Luxembourg, Luxembourg, L-1359, Luxembourg, \EMAIL{nils@loehndorf.com}}

\AUTHOR{David Wozabal}
\AFF{Department of Operations Analytics, Vrije Universiteit Amsterdam School of Business and Economics, De Boelelaan 1105, 1081 HV Amsterdam, Netherlands, \EMAIL{d.wozabal@vu.nl}}
} 

\ABSTRACT{
Corporate renewable power purchase agreements (PPAs) are long-term contracts that enable companies to source renewable energy without having to develop and operate their own capacities. Typically, producers and consumers agree on a fixed per-unit price at which power is purchased. The value of the PPA to the buyer depends on the so called capture price defined as the difference between this fixed price and the market value of the produced volume during the duration of the contract. To model the capture price, practitioners often use either fundamental or statistical approaches to model future market prices, which both have their inherent limitations. We propose a new approach that blends the logic of fundamental electricity market models with statistical learning techniques. In particular, we use regularized inverse optimization in a quadratic fundamental bottom-up model of the power market to estimate the marginal costs of different technologies as a parametric function of exogenous factors. We compare the out-of-sample performance in forecasting the capture price using market data from three European countries and demonstrate that our approach outperforms established statistical learning benchmarks. We then discuss the case of a photovoltaic plant in Spain to illustrate how to use the model to value a PPA from the buyer's perspective.
}


\maketitle
\section{Introduction} \label{sec:intro}
The manufacturing sector remains a significant contributor to carbon emissions, accounting for about a quarter of energy-related emissions, largely due to its continued reliance on fossil fuels like coal and gas~\citep{iea2023industry}. To meet the EU's climate targets and the sustainability goals in their supply chains, European manufacturers are heavily investing in low-carbon production techniques, but also participate directly in the development of renewable energy projects. 

Given the inherent limitations in producing the required amount of renewable electricity on site, large industrial consumers often utilize power purchase agreements (PPAs) with renewable energy producers. PPAs are long-term agreements between consumers and producers of electricity and are becoming an increasingly popular instrument to reach sustainability targets of energy intensive industries: Corporations bought a record $31.1$ GW of clean electricity through PPAs in 2021, up nearly 24\% from the previous year’s record of $25.1$ GW \citep{bnef}.

Typically, a producer and a consumer agree on a fixed price at which power is purchased over a horizon of $5$ to $20$ years. The long-term stability of revenue streams helps producers to enhance creditworthiness and acquire the necessary capital for their investment as well as large consumers to meet their sustainability goals. However, while consumers benefit from price stability as well, they also face exposure to the risk of low market prices due to cannibalization effects or renewable power \citep{hirth2013market}. Large industrial buyers, such as data center operators, chemical companies, and steel makers, for whom energy constitutes a substantial share of their total cost, therefore find themselves in urgent need of models to evaluate the PPAs offered to them. 

Valuation of PPAs is challenging and the lack of corresponding knowledge and models is a main barrier for PPA adoption \citep{pwc2016corporate,ghiassi2021making}. The value of a PPA depends on the difference between the fixed electricity price the consumer pays to the producer and the future market price of power. The latter is modeled by the so-called capture price defined as the volume-weighted average price with volumes given by the renewable production quantities contracted under the PPA. Calculating the PPAs capture price is therefore central to its valuation.

The literature on valuation of PPAs is sparse. \cite{pircalabu2017joint} and \cite{tranberg2020managing} propose time series models of the joint process of electricity prices and renewable generation to calculate the profit distribution of PPAs. However, their models are not suitable for long-term valuation, as they do not take fundamental drivers of power prices into account. \cite{trivella2023meeting} model the problem of meeting corporate renewable power targets through the purchase of PPAs as a dynamic portfolio optimization problem. However, their focus lies on solving the underlying Markov decision process and not on pricing individual PPAs or calculating capture prices. 

An alternative approach often found in practice is to compare the price of a PPA to levelized cost of energy (LCOE) of the corresponding asset \citep{bruck2018levelized, mendicino2019corporate, bruck2021pricing, lindahl2022economic, gabrielli2022mitigating, gabrielli2022storage}. However, the LCOE can merely be viewed as a lower bound of the PPA value which would only be exact in a perfectly competitive market where producers merely get reimbursed their long-term cost.

Correctly calculating the capture price entails modeling market equilibria. These equilibria are affected by power consumption patterns and the availability of certain production technologies, including storage and transport capacities, which will likely fundamentally change in the foreseeable future. Valuation of a PPA therefore entails modeling electricity prices over long time periods with changing power market fundamentals.

The literature on models of electricity prices can be broadly split into three groups:  statistical and machine learning models \citep{nogales2002forecasting,contreras2003arima,conejo2005day,zhang2005neural, taylor2010triple}, fundamental price models \citep{boogert2008supply,howison2009stochastic,aid2013structural,carmona2013electricity}, and stochastic models \citep{schwartz1997stochastic,schwartz2000short,lucia2002electricity,geman2005soybean,carmona2013electricity,secomandi2014optimal,heath2019macroeconomic}.

Statistical approaches typically estimate a functional relationship between exogenous factors and market prices based on historical data. While these models are often successful in short-term forecasting, they are not suitable for long horizons due to their inability in handling distribution shifts, for example, sudden drops in demand, changes in the supply side, or extreme price regimes, which can be expected to occur over the lifetime of a PPA~\citep{weron2014electricity}. 

Stochastic models are typically used for pricing and risk management but are structurally limited when it comes to long-term forecasting as parameters are calibrated to observed prices such as forward curves and do not consider the evolution of fundamental factors.

Fundamental models, on the other hand, capture the physical realities of the power system and the complex interactions between various production technologies and are theoretically capable of predicting market prices for any hypothetical scenario. However, their predictions often perform poorly, as it is hard to accurately model the cost structure along with all operational characteristics of the power sector. Hence, these models require expert calibration and are often outperformed by statistical models in short-term forecasting~\citep{weron2014electricity, pape2016fundamentals}. 

In this article, which grew out of a cooperation with \emph{ArcelorMittal}, the world's second largest steel producer, we combine the best of statistical methods and fundamental models. 
In order to build a fundamental model that accurately explains electricity price formation over a long period of time, it is essential to understand cost and operational characteristics of the generation technologies. While historical prices and production quantities for different technologies are publicly available, marginal costs are typically confidential and difficult to elicit. To overcome this problem, we propose a new approach to model electricity prices that is based on inverse optimization. 

Seminal work on inverse optimization can be found in \cite{Ahuja2001} who investigate the problem of estimating objective function coefficients for general linear programs. \cite{iyengar2005inverse} and \cite{zhang2010inverse} extended this approach to conic programs and \cite{schaefer2009inverse} and \cite{wang2009cutting} to integer programs. \cite{aswani2018inverse} treat the case when data is noisy. Our approach aligns with prior work in inverse optimization, particularly the framework presented by~\cite{zhang2010augmented}, who consider the inverse problem of a quadratic program. For a recent and comprehensive review of both the methodological and application-oriented literature in this field, see \cite{chan2023inverse}. 

In this paper, we use the inverse of a fundamental electricity market model to estimate how external factors influence marginal cost of production. In particular, we use historical equilibrium production and aim to find marginal production prices that yield the observed market equilibria when used in the fundamental model.

However, as the fundamental model is merely an approximation of the underlying market realities, it is not always possible to find marginal costs that exactly replicate equilibria. Instead, we minimize a measure of suboptimality \citep{keshavarz2011imputing,chan2014generalized,bertsimas2015data,mohajerin2018data,aswani2018inverse,chan2019inverse}. To that end, we seek to explain marginal cost of plants by a range of features and take the perspective of a regression problem minimizing the empirical loss induced by the forecast errors. Additionally, and similar to LASSO regression \citep{tibshirani1996regression}, we introduce a regularization term in the objective that avoids overfitting the insample data. We model production cost per technology and posit a quadratic relationship between produced quantity and cost. As we will show, the quadratic structure allows for the emulation of unit-specific costs within a group of power plants of the same technology, improving model performance.

The parameters estimated from the inverse problem are incorporated into a so-called forward problem, where production quantities are decision variables and marginal cost of each technology is predicted based on available exogenous factors using the estimated coefficients of the inverse problem. This approach combines the strength of fundamental models to explain prices in regimes that are not contained in the data with the advantages of machine learning that is able to \emph{learn} from historical data.

In a numerical study, we compare our model with two pure machine learning benchmarks --- simple LASSO regression \citep{tibshirani1996regression} and XGBoost \citep{chen2016xgboost}, a popular gradient boosting approach. The inverse optimization model clearly and consistently outperforms both benchmarks out-of-sample across multiple time horizons and markets. Specifically, we demonstrate that the model can handle distribution shifts, which occur when test data follows a different distribution than training data as was the case during the COVID pandemic or the 2021-22 gas price hike that followed the onset of the Russo-Ukrainian war. 

Finally, we describe how the model supports analysts and buyers in price negotiations based on field data from a large European steel maker. We explain how buyers tasked with evaluating a PPA can use the forward model to calculate the capture price based on analysts' custom scenarios of the possible evolution of fundamental factors. The resulting scenarios then provide buyers with additional information when negotiating the price of PPAs with producers. 

In summary, our work contributes to three strands of literature at the interface of sustainable supply chain management, operations research, and energy economics: 
\begin{enumerate}
    \item We contribute to research on valuation of PPAs by providing buyers with a model to calculate capture prices over the lifetime of a PPA while accounting for changes in the underlying market fundamentals.
    
    \item We make a methodological contribution to inverse optimization by addressing the challenge of handling noisy model parameters using regularized regression within the context of an inverse quadratic programming problem.
    
    \item Ultimately, our research contributes to the literature on electricity price models, being the first application of inverse optimization to fundamental price models.
\end{enumerate}

The remainder of the paper is organized as follows: Section \ref{section:PPA} defines the valuation problem in mathematical terms and introduces the forward model and its inverse. Section \ref{section:Results} summarizes the result of extensive backtests based on market data from Germany, France, and Spain, and contains a comparison of the inverse model with benchmarks based on statistical learning. Section \ref{section:CaseStudy} illustrates how buyers can use the model in PPA procurement using the example of a PPA for a solar power plant in Spain. A discussion and an outlook on future work is given in Section \ref{section:Conclusion}.

\section{Model} \label{section:PPA}
In this section, we describe the pricing mechanism for PPAs and review some basic assumptions in Section \ref{ssec:pricing_principle}. In Section \ref{ssec:Forward_problem}, we introduce a fundamental bottom-up market model that yields equilibrium prices as forecasts for electricity prices. Finally, we propose a machine learning based inverse optimization approach to estimate the parameters of the model from observed market prices and fundamental factors in Section \ref{ssec:inverse_opt}.

\subsection{Valuing Renewable PPAs} \label{ssec:pricing_principle}
For our valuation approach, we assume that the start of delivery of electricity from the PPA is at $t=0$, the contract ends at $t=T$, and denote by $\Tc = \{1, \dots, T\}$ the set of time periods. The renewable power source produces an amount $q_t$ of electricity for $t\in \Tc$ which is delivered to the buyer of the PPA for a fixed price $p$ per MWh. 

The value of owning a PPA, $\pi$, has two components: the value of the electricity $\pi^E$ that is delivered through the contract and the added value from the fact that the power is generated from a renewable source of electricity $\pi^G$. We can write this as a function of $p$ as follows
\begin{align}
    \pi(p) &= \underbrace{\sum_{t\in \Tc} \rho_t q_t (p_t^E-p)}_{\pi^E} + \underbrace{\sum_{t\in \Tc} \rho_t q_t p_t^G}_{\pi^G}, \label{eq:PPA_profit}
\end{align}
where $\rho_t$ is the discount factor, $p_t^E$ is the market price for conventionally produced electricity at time $t$, and $p_t^G>0$ is the per MWh markup that represents the additional value of renewable power to the buyer of the PPA. Hence, $\pi(p)$ can be viewed as the value of an electricity future on a fluctuating amount of green electricity $(q_t)_{t\in \Tc}$.

The markup, $p_t^G$, is a premium for sourcing renewable energies that is driven by idiosyncratic factors such as tax credits, reduced financing costs, or the strategic and marketing benefits from reducing the firm's carbon footprint. Viewing $p_t^G$ purely as a sustainability premium, it can also be interpreted as the market price for certificates of origin. 

Note that by signing a PPA, both parties completely hedge their price risks, since $q_t$ is traded for a fixed price $p$ instead of a random future market price $p_t^E$. From a seller's perspective, PPAs are thus similar to fixed feed-in tariffs that were used in many countries to incentivize investments in renewable energy.

We propose a pricing mechanism that yields a maximum price that a buyer would be willing to pay for a PPA by choosing $p$ such that $\pi(p)$ equals zero, i.e.,
\begin{align} \label{eq:PPA_price_risk_neutral}
    p = \frac{\sum_{t=1}^T \rho_t q_tp_t^E + \sum_{t=1}^T \rho_t q_tp_t^G}{\sum_{t=1}^T \rho_t q_t}.
\end{align}
Clearly, the above formula for $p$ neglects the risk preferences as well as the bargaining power of the buyer and the seller and therefore cannot fully reflect the prices paid for PPAs in the market. However, it is an essential piece of information for the buyer when bidding for PPAs and in price negotiations.

Note that if we neglect discounting, i.e., set $\rho_t = 1$ for all $t$, and do not assign any additional value to green electricity, i.e., set $p_t^G = 0$, \eqref{eq:PPA_price_risk_neutral} further simplifies to 
\begin{align} \label{eq:capture_price}
    p \equiv \frac{\sum_{t=1}^T q_tp_t^E}{\sum_{t=1}^T q_t},
\end{align}
which is commonly referred to as the average \emph{capture price} of the PPA.

The aim of this paper is to find a $p$ that solves \eqref{eq:PPA_price_risk_neutral}. In order to do so, we require $p_t^E$ and $p_t^G$ as well as the quantities $q_t$ as inputs. Since $p_t^G$ is elusive and ultimately depends on the buyer's individual valuation of sustainability, this quantity cannot be estimated from a general model. The produced quantities $q_t$ depend on weather conditions and can be assumed to follow a stationary seasonal pattern and while production is hard to forecast for a specific point in time $t$, cumulative production over several years can be predicted with high accuracy. We therefore argue that the most difficult part in pricing a PPA is forecasting future electricity prices $p_t^E$. In the rest of this section, we therefore focus on this problem and derive a model that is able to produce consistent price forecasts from fundamental information about the evolution of the electricity sector.

\subsection{Fundamental Electricity Price Model} \label{ssec:Forward_problem}
As argued above, in order to evaluate the PPA, we need to forecast the long-term development of prices and produced quantities over the duration of the contract. Power prices are highly non-stationary and can therefore only be modeled for short time horizons using statistical models. Likewise, price models based on forward price quotations are limited to block products (base, peak) and by a lack of liquidity of traded long-term electricity futures.

In the following, we focus on modeling the price for electricity $p_t^E$ in order to get a forecast for $\pi_t^E$ and set $\pi = \pi^E$. We therefore neglect the added benefit of carbon neutral production and consequently obtain lower bounds on the price of PPAs that need to be increased by the idiosyncratic willingness of the buyer to pay a premium for \emph{green electricity}. 

Of course, in reality there is not \emph{one} price for electricity as there are several staggered futures markets that trade contracts for the same delivery period. Hence, we think of the price as the day-ahead price for electricity, since most electricity markets feature a liquid day-ahead market which is typically the \emph{first} market that trades products in hourly resolution. Furthermore, the day-ahead market is the spot market that generally serves as a reference for electricity prices. 

To calibrate long-term price models, we employ a fundamental, bottom-up modeling approach that is able to accommodate anticipated structural changes in the electricity sector such as expansion of renewable capacities, the phase out of conventional generation, electricity demand, and changing prices for inputs such as fossil fuels and emission allowances.

\begin{figure}[ht]
    \centering
    \includegraphics[width = \textwidth]{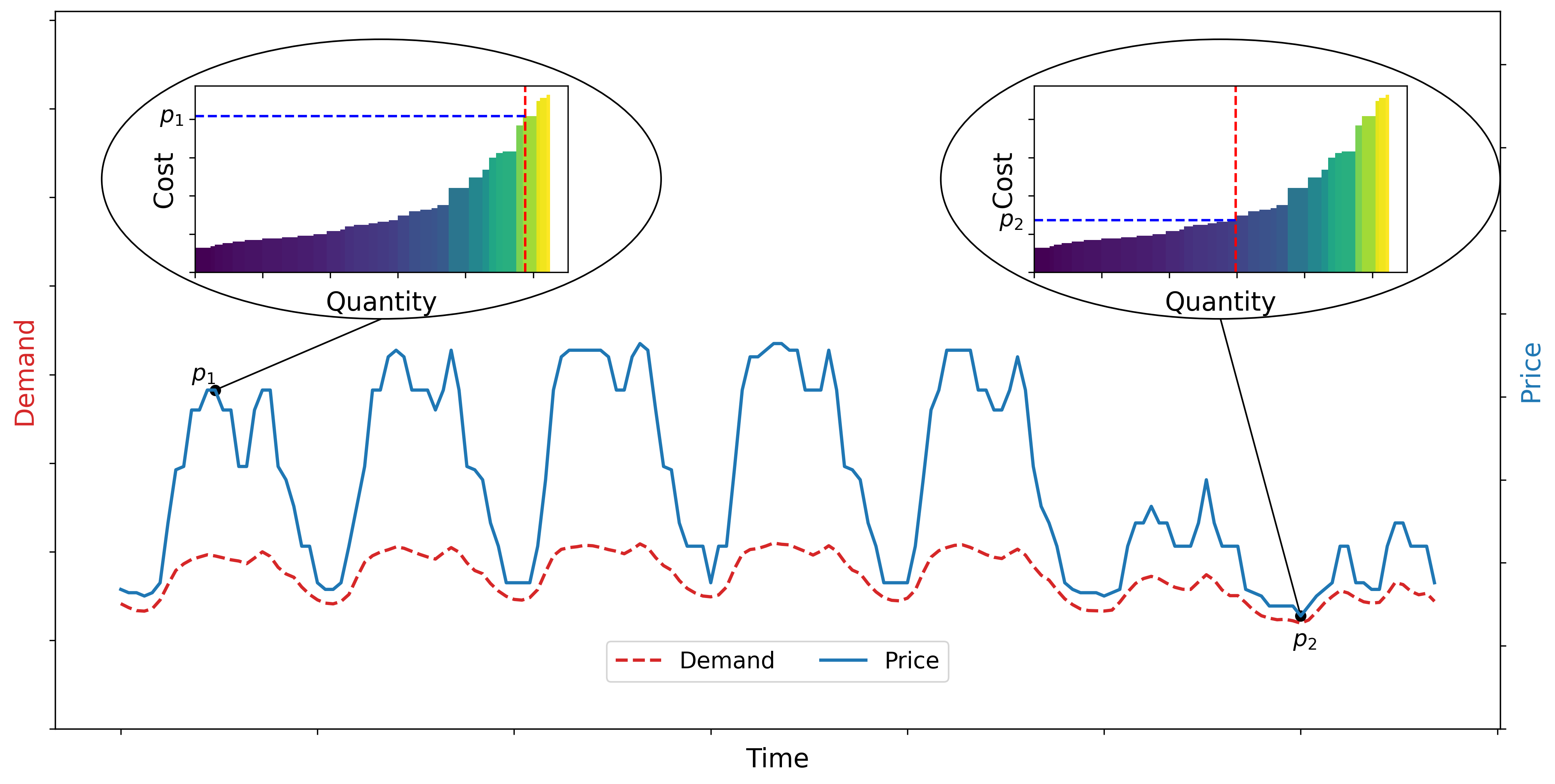}
    \caption{\label{fig:merit_order} Stylized depiction of the effect of demand on prices in a simple merit order model for one week in hourly resolution. The encircled left plot shows the situation in a high demand hour around noon on Monday, whereas the right one depicts a low demand on a Sunday morning. The bars in the smaller graphs represent the individual power plants ordered according to their marginal cost of production with the width of the bars corresponding to capacities and the height to marginal cost. In aggregate the bars represent the inverse supply function in the market. The inelastic demand is represented by the dotted vertical lines.}
\end{figure}

The basic idea of our approach is rooted in the merit order model \citep[also known as the bid stack model, see][]{wolyniec2003energy} that defines electricity prices as the intersection of the aggregate inverse supply curve with a completely inelastic demand for electricity. See Figure~\ref{fig:merit_order} for an illustration of how the merit order determines the price for different levels of demand. In particular, if $c_i$ is the short-run marginal production cost of supplier $i$ with capacity $\bar X_i$, and the demand for electricity is $d$, then the price according to the merit order model is 
$$ p = \min_i\left\{c_i: \sum_{j=1}^i \bar X_i \geq d \right\},$$
assuming that marginal costs are ordered, i.e., $c_i \leq c_{i+1}$. The resulting price is the welfare maximizing market clearing price in the static setting above.

Apart from the assumption of inelastic demand, the accuracy of the model is predicated on several assumptions that guarantee the validity of the first fundamental theorem of welfare economics and ensure that market outcomes yield a welfare optimal allocation. In particular, the model implicitly assumes perfect information, perfect competition, no non-convexities, and deterministic planning by all market participants, i.e., that risk preferences do not play a role. See \cite{BJORNDAL2008768} and \cite{ralph2015,ferris2021} for models that relax the assumptions on non-convexities and deterministic planning, respectively.

Another problem with the basic merit order model is that it is static, meaning it models a single market clearing without regard to market results in earlier or later periods. Due to the scarcity of electricity storage and the need to constantly balance load and production, there are many market clearings for one day~--~typically one market for every hour or even quarter-hour. However, because of the dynamic restrictions imposed by ramping cost and ramping limits as well as the dynamics imposed by electricity storage, the cost structures and therefore the decisions of producers can only be accurately captured in a dynamic model that explicitly links time periods.

To accommodate some of these dynamic features, let us move from the implicit static welfare optimization model to an explicit dynamic welfare optimization, which in the case of inelastic demand takes the form of a cost minimization problem of a central planner. While the dynamic model captures more resource constraints, we keep the assumption of perfect competition and inelastic demand. 

The aim of the central planner is to fulfill the energy demand $d_t$ in every time period $t \in \Tc$ at minimal cost by dispatching available production technologies across time in an optimal way. To that end, we consider a set of technologies $\Ic$ and, for each technology $i\in \Ic$, we define production quantities $x_{it}$ and quadratic production cost $C_{it}(x_{it}) = c^1_{it}x_{it} + c^2_{it}x_{it}^2$ at time $t$ as well as the overall power capacity $\bar X_{it}$. Due to the non-linearity of fuel efficiency, transmission losses, and the cost of reactive power, the choice of quadratic cost functions is standard in the economic dispatch literature \citep[e.g.][]{Cain2004, Oren2008, wood2013power}. Furthermore, quadratic cost allow us to capture the fact that within each technology group cheaper power plants are dispatched before more expensive ones.

We assume that there is storage with an energy capacity of $\bar S_t$ (in MWh), a power capacity of $\bar Y_t$ (in MW), and a (symmetric) efficiency $\eta \in (0,1)$. We denote the storage level by $s_t$, and the charging and withdrawal by $y_t^+$ and $y_t^-$ (in MWh), respectively. We further assume that conventional plants have a limited ramp-up capacity of $R_{it}^+$ (in MW) and a ramp-down capacity of $R_{it}^-$ (in MW). To model cost associated with ramping, we let $r_{it}^+$ and $r_{it}^-$ denote the ramp-up and ramp-down decision for technology $i$ at time $t$ and $k_{it}$ to be the cost of per MW ramp-up of technology $i$ at time $t$.
The decision problem of the planner can then be written as
\begin{subequations} \label{eq:forward_problem}
    \begin{align}
        \min \quad & \sum_{i \in \mathcal{I}}\sum_{t \in \mathcal{T}} C_{it}(x_{it}) + k_{it}r^+_{it} & \label{eq:forward_obj} \\ 
        \text{s.t.} \quad &  x_{it} \leq \bar X_{it} & \forall i\in \mathcal{I},\; \forall t\in    \mathcal{T} &\:\:\:\:\:\: [\overline \alpha_{it}] & \label{eq:capacity_gen_high} \\ 
        & 0 \leq x_{it} & \forall i\in \mathcal{I},\; \forall t\in    \mathcal{T} &\:\:\:\:\:\: [\underline \alpha_{it}] & \label{eq:capacity_gen_low} \\ 
        &\sum_{i \in \mathcal{I}}x_{it} + y_{t}^- - y_t^+ = d_{t} & \forall t \in \mathcal{T}& \:\:\:\:\:\:[p^E_{t}] & \label{eq:demand_constraint} \\ 
        & s_{t} = s_{t-1} + \eta y_{t}^+ - \eta^{-1} y_{t}^- & \forall t\in \mathcal{T}& \:\:\:\:\:\:[\pi_{t}] & \label{eq:storage_balance} \\ 
        & s_{t} \leq \bar S_{t} & \forall t\in \mathcal{T}& \:\:\:\:\:\:[\overline \beta_{t}] & \label{eq:energy_capacity_storage_high} \\ 
        & 0 \leq s_{t} & \forall t\in \mathcal{T}& \:\:\:\:\:\:[\underline \beta_{t}] & \label{eq:energy_capacity_storage_low} \\ 
        & y_t^+ \leq \bar Y_{t}^+ & \forall t\in \mathcal{T}& \:\:\:\:\:\:[\overline \gamma_{t}^+] & \label{eq:power_capacity_storage_discharge_high} \\ 
        & 0 \leq y_t^+ & \forall t\in \mathcal{T}& \:\:\:\:\:\:[\underline \gamma_{t}^+] & \label{eq:power_capacity_storage_discharge_low} \\ 
        & y_t^- \leq \bar Y_{t}^- & \forall t\in \mathcal{T}& \:\:\:\:\:\:[\overline \gamma_{t}^-] & \label{eq:power_capacity_storage_charge_high} \\ 
        & 0 \leq y_t^-  & \forall t\in \mathcal{T}& \:\:\:\:\:\:[\underline \gamma_{t}^-] &\label{eq:power_capacity_storage_charge_low}\\
        & x_{it} - x_{i,t-1} = r_{it}^+ - r_{it}^- & \forall i\in \mathcal{I},\; \forall t\in    \mathcal{T}& \:\:\:\:\:\:[\mu_{it}] & \label{eq:SSR1} \\ 
        & r_{it}^+ \leq R_{it}^+ & \forall i\in \mathcal{I},\; \forall t\in    \mathcal{T}& \:\:\:\:\:\:[\overline \delta_{it}] & \label{eq:SSR2} \\ 
        & 0 \leq r_{it}^+ & \forall i\in \mathcal{I},\; \forall t\in    \mathcal{T}& \:\:\:\:\:\:[\underline \delta_{it}] & \label{eq:SSR3} \\ 
        & r_{it}^- \leq R_{it}^- & \forall i\in \mathcal{I},\; \forall t\in    \mathcal{T}& \:\:\:\:\:\:[\overline \theta_{it}] & \label{eq:SSR4} \\ 
        & 0 \leq r_{it}^- & \forall i\in \mathcal{I},\; \forall t\in    \mathcal{T}& \:\:\:\:\:\:[\underline \theta_{it}] & \label{eq:SSR5} 
    \end{align}
\end{subequations}
where $s_0$ is the initial storage state. The constraints \eqref{eq:capacity_gen_high}-\eqref{eq:capacity_gen_low} enforce capacity restrictions for each technology, \eqref{eq:demand_constraint} makes sure that the demand is met, \eqref{eq:storage_balance} models storage dynamics, \eqref{eq:energy_capacity_storage_high}-\eqref{eq:energy_capacity_storage_low} enforce storage capacity limits,  \eqref{eq:power_capacity_storage_discharge_high}-\eqref{eq:power_capacity_storage_charge_high} restrict the power capacity of storage, and the constraints in \eqref{eq:SSR1}--\eqref{eq:SSR5} model ramping limits.

Note that for later reference, the dual multipliers of the constraints are reported in square brackets. In particular, the dual multipliers of \eqref{eq:demand_constraint} are the shadow prices of a marginal increase in demand, which are the market prices in a competitive market that fulfills all the above-mentioned assumptions. 

Because the fundamental market mechanism is modeled by explicitly describing all components of the system in a bottom-up manner, we refer to models of type \eqref{eq:forward_problem} as fundamental bottom-up models or fundamental models in short. This structural approach to energy systems modeling is very common in the analysis of existing systems, in planning future energy systems, and generally for the analysis of the effect of policy decisions \citep[e.g.,][]{weigt2008price,hirth2013market,lopion2018review}. 

The results of the market model in \eqref{eq:forward_problem} depends on several parameters, including the demand for electricity, the capacities of the different technologies including storage, as well as the costs for production inputs. In conventional fundamental models these parameters are entirely exogenous and specified by the modeler. This entails collecting data on fuel costs, plant efficiencies, must-run restrictions, downtimes, and many other parameters for every single power plant.

However, it is generally hard to accurately capture certain aspects of operational decision making in electricity generation. In particular, it is difficult to estimate real efficiencies of power plants, dynamic constraints connected to ramping, cost of equipment wear, as well as operational philosophies of plant operators.

We therefore take a novel approach by estimating some of the parameters using inverse optimization as discussed in the next section. In particular, we observe overall electricity demand as well as production quantities and capacities for every technology. Based on this information, we infer the marginal prices and ramping cost of the technologies, assuming that the model \eqref{eq:forward_problem} governs the price dynamics. We express the marginal cost of each technology as a function of exogenous features. The resulting model is a hybrid between classic fundamental bottom-up modeling and machine learning. This enables us to better align the model with observed market outcomes and capture certain properties of prices, such as spikes, which are otherwise very hard to explain in purely fundamental models.

\subsection{Inverse Optimization} \label{ssec:inverse_opt}
As discussed above, in order to calibrate \eqref{eq:forward_problem}, we use an inverse optimization approach to estimate a model for $C_{it}$ which leads to observed decisions $x_{it}$, $s_t$, $y_t^+$, $y_t^-$, $r_{i,t}^+$, and $r_{i,t}^-$. Generally, in inverse optimization the values of certain model parameters are calibrated by solving a \emph{backward optimization} problem to fit the observed optimal decisions of a \emph{forward optimization} problem, in our case \eqref{eq:forward_problem}.

We require the following result for generic linearly constrained quadratic optimization problems, which in particular covers problem \eqref{eq:forward_problem}. The result follows as a special case from \cite{zhang2010inverse} but we nevertheless provide an elementary proof for the convenience of the reader.
\begin{proposition} \label{prop:inverse_quadratic}
    Consider the following problem
    \begin{subequations} \label{eq:generic_quadratic}
        \begin{align}
            \min_x  \quad  & \sum\limits_{i=1}^n c^1_i x_i + \sum_{i=1}^n c^2_i x_i^2 \\
            \text{s.t.} \quad   & Ax \leq b & \:\: [\lambda] \\
                                & x \leq \overline X & [\overline \mu]\\
                                & \underline{X} \leq x & [\underline \mu]
        \end{align}
    \end{subequations}
    where $A = (a_1, \dots, a_m)^\top \in \Rd^{m\times n}$ with $a_i \in \Rd^n$, i.e., there are $m$ linear inequality constraints. Define the sets $\Uc(\lambda) = \{j: \langle a_j, x \rangle < b_j, \; 1 \leq j \leq m \}$, $\Uc(\underline \mu) = \{i \in \Ic: x_i < \overline X_i \}$, and $\Uc(\overline \mu) = \{i \in \Ic: \underline X_i < x_i\}$ as the set of indices where the respective constraints are not binding.
    
    Given an optimal solution $x^*$ for \eqref{eq:generic_quadratic}, the objective value coefficients $c$ and $q$ and the dual variables $\lambda$, $\overline \mu$, $\underline \mu$ have to fulfill the following system of equations
    \begin{align}
        \lambda \leq 0, \; \underline \mu \leq 0, \; \overline \mu \leq 0 & \label{eq:dual_feasibility}\\
        \lambda_j = 0 \quad \forall j \in \Uc(\lambda), \quad \overline \mu_i = 0 \quad \forall i \in \Uc(\overline \mu), \quad & \underline \mu_i = 0 \quad \forall i \in \Uc(\underline \mu) \label{eq:slackness}\\
        c^1_i + 2c^2_ix_i^* - \sum_{j=1}^m A_{ji} \lambda_j + \underline \mu_i - \overline \mu_i = 0, &\quad \forall i = 1, \dots, n. \label{eq:lagrangian}
    \end{align}
\end{proposition}
\proof{Proof.}
    Since the above problem is a convex optimization problem, the KKT conditions are necessary and sufficient for an optimal solution. 
    
    Definingn $\langle \cdot, \cdot \rangle$ as the inner product, the Lagrangian of the problem is
    \begin{align}
        \Lc(x, \lambda, \underline \mu, \overline \mu) = \sum\limits_{i=1}^n c^1_i x_i + \sum_{i=1}^n c^2_i x_i^2 + \langle \lambda,  b- Ax \rangle + \langle \underline \mu, x - \underline X \rangle  + \langle \overline \mu, \overline X - x \rangle.
    \end{align}
    Taking the derivative with respect to $x_i$ yields
    \begin{align}
        \frac{\partial}{\partial x_i} \Lc(x, \lambda, \underline \mu, \overline \mu) = c^1_i + 2c^2_ix_i - \sum_{j=1}^m A_{ji} \lambda_j + \underline \mu_i - \overline \mu_i.
    \end{align}
    Hence, conditions \eqref{eq:lagrangian} ensures that the gradient of the Lagrangian at $x^*$ is zero, while \eqref{eq:dual_feasibility} ensures dual feasibility of the solution, and primal feasibility is fulfilled by the assumptions on $x^*$. Finally, the complementary slackness conditions hold because of \eqref{eq:slackness}. \hfill \Halmos
\endproof

In order to learn a model for $C_{it}$ in \eqref{eq:forward_problem} based on decisions $x_{it}$, $s_t$, $y_t^+$, $y_t^-$, $r_{i,t}^+$, and $r_{i,t}^-$, we define the sets
\begin{align}
    \Uc(\overline \alpha) &= \{(i,t) \in \Ic \times \Tc: x_{it} < \bar X_{it} \}, \; &\Uc(\underline \alpha) &= \{(i,t) \in \Ic \times \Tc: 0 < x_{it} \} \\
    \Uc(\overline \beta) &= \{t \in \Tc: s_t < \bar S_t\}, \; &\Uc(\underline \beta) &= \{t \in \Tc: 0<s_t\} \\
    \Uc(\overline y^+) &= \{t \in \Tc: y_t^+ < \bar Y_t^+\}, \; &\Uc(\underline y^+) &= \{t \in \Tc: 0<y_t^+\} \\
    \Uc(\overline y^-) &= \{t \in \Tc: y_t^- < \bar Y_t^-\}, \; &\Uc(\underline y^-) &= \{t \in \Tc: 0 < y_t^-\} \\
    \Uc(\overline \delta) &= \{(i,t) \in \Ic \times \Tc: r_{i,t}^+ < \bar R_{i,t}^+\}, \; &\Uc(\underline \delta) &= \{(i,t) \in \Ic \times \Tc: 0<r_{i,t}^+\} \\
    \Uc(\overline \theta) &= \{(i,t) \in \Ic \times \Tc: r_{i,t}^- < \bar R_{i,t}^-\}, \; &\Uc(\underline \theta) &= \{(i,t) \in \Ic \times \Tc: 0<r_{i,t}^-\} 
\end{align}
of time periods where the respective dual variables are zero due to complementary slackness. For example, $\overline \alpha_{it} = 0$ for all $t \in \Uc_i(\overline \alpha)$ because the power limit for technology $i$ is not binding at time $t$ in the optimal solution. In conventional inverse optimization, we would choose the cost coefficients $c_{it}^1$, $c_{it}^2$, and $k_{it}$ such that the distance to some prior guess for the parameter is minimized while respecting all available information on the primal and dual solution \citep[see, e.g.,][]{Ahuja2001}. 

Because we do not want to choose $C_{it}$ and $k_{it}$ for every $i \in \Ic$ and $t \in \Tc$ separately but rather want to learn a model that predicts $C_{it}$ from the values of observable exogenous variables, we fit a regression that explains $C_{it}$ and $k_{it}$ as a function of features \(Z_{it}^1 \in \Rd^{n_i^1}\), \(Z_{it}^2\in \Rd^{n_i^2}\), and \(Z_{it}^3\in \Rd^{n_i^3}\) as
\begin{equation} \label{eq:regression}
\begin{aligned}
    c^1_{it} &= \langle Z^1_{it}, b_i^1 \rangle + \varepsilon_{it}^1 \\
    c^2_{it} &= \langle Z^2_{it}, b_i^2 \rangle + \varepsilon_{it}^2 \\
    k_{it} &= \langle Z^3_{it}, b_i^3 \rangle + \varepsilon_{it}^3,
\end{aligned}
\end{equation}
where $b_i^j \in \Rd^{n_i^j}$ are vectors of coefficients that model the linear relation between $Z_{it}^1$, $Z_{it}^2$, and $Z_{it}^3$ and $c^1_{it}$, $c^2_{it}$, and $k_{it}$, respectively and $\langle \cdot, \cdot, \rangle$ are the inner products in the corresponding spaces. Note that the equations in \eqref{eq:regression} resemble linear regressions with error terms $\varepsilon_{it}^j$ accounting for the fact that marginal prices cannot be precisely modeled using features $Z_{it}^j$, thereby ensuring that the problem remains feasible.

Thus, instead of minimizing the distance of $C_{it}$ and $k_{it}$ to some initial guess we seek to minimize the error in \eqref{eq:regression} expressed by the variances of $\varepsilon_{it}$. Since we potentially have many features, i.e., $n_i$ is large, and we do not want to overfit the model in order to guarantee a good out-of-sample performance, we additionally introduce an L$^\text{1}$-penalty for the  magnitude of $b_i$. Furthermore, we introduce weights $w_t$ for the errors in period $t$ in order to put the focus of the model on time periods that are more relevant for the pricing of a particular PPA. We therefore end up with the following objective function for the inverse optimization problem
\begin{align*}
    \mathlarger{\sum}\limits_{i,t}w_t\left((c^1_{it}-\langle Z_{it}, b_i^1 \rangle)^2 + (c^2_{it}-\langle Z_{it}, b^2_i \rangle)^2 + (k_{it}-\langle Z_{it}, b''_i \rangle)^2\right) + \mathlarger{\sum}\limits_{ij} \lambda^j_i ||b_i^j||_1
\end{align*}
In the numerical results section, we set $w_t$ to the expected capacity factors in $t$ of the technology we want to price a PPA for. In this case, the errors can be interpreted as the errors in the forecast of the capture price. If $w_t\equiv 1$ is chosen and all $t \in \Tc$ get the same weight, then the model tries to forecast the base price as good as possible.

In order to derive the constraints of the inverse problem to \eqref{eq:forward_problem}, we note that the Lagrangian of the problem can be written as
\begin{equation}
\begin{split}
\Lc &= \sum_{i,t} \left( c^1_{it} x_{it} + c^2_{it} x_{it}^2 + \overline{\alpha}_{it} (\overline{X}_{it} - x_{it}) + \underline{\alpha}_{it} x_{it} + \mu_{it}(r_{it}^+ - r_{it}^- - x_{it}+x_{i,t-1}) \right. \\
&\quad + \overline{\delta}_{it}(R^+_{it}-r^+_{it}) + \underline{\delta}_{it} r_{it}^+  + \overline{\theta}_{it}(R^-_{it}-r_{it}^-) +\underline{\theta}_{it}r^-_{it}+r_{it}^+ k_{it} \\
& \quad + \left. \sum_t \left( p_t^E(d_t - \sum_{i} x_{it} - y_t^- + y_t^+) + \pi_t (s_t - s_{t-1} -\eta y_t^+ + \eta^{-1}y_t^-) \right. \right. \\
&\quad\quad + \overline{\beta}_t (\overline{S}_t - s_t) + \underline{\beta}_t s_t + \overline{\gamma}_t^+ (\overline{Y}_t^+ -y_t^+) + \underline{\gamma}_t^+y_t^+ \\
&\quad\quad + \left. \overline{\gamma}_t^- (\overline{Y}_t^- - y_t^-) + \underline{\gamma}_t^- y_t^- \right).
\end{split}
\end{equation}

To make use of Proposition \ref{prop:inverse_quadratic}, we take the derivative of the Lagrangian with respect to the primal decision variables as follows,
\begin{align}
    \frac{\partial}{\partial x_{it}} \Lc &= c^1_{it} + 2c^2_{it}x_{it} - \overline \alpha_{it} + \underline \alpha_{it} - p_t^E -\mu_{it}+\mu_{i,t+1} \\
    \frac{\partial}{\partial y_t^+} \Lc &= p_t^E - \eta \pi_t - \overline \gamma_{t}^+ + \underline \gamma_{t}^+ \\ 
    \frac{\partial}{\partial y_t^-} \Lc &= -p_t^E + \eta^{-1} \pi_t - \overline \gamma_{t}^- + \underline \gamma_{t}^- \\ 
    \frac{\partial}{\partial s_t} \Lc &= \pi_t - \pi_{t+1} - \overline \beta_t + \underline \beta_t\\
    \frac{\partial}{\partial r_{it}^+} \Lc &= \mu_{it} - \overline \delta_{it} + \underline \delta_{it} + k_{it}\\
    \frac{\partial}{\partial r_{it}^-} \Lc &= -\mu_{it} - \overline \theta_{it} + \underline \theta_{it} 
\end{align}
In summary, for a given observed optimal solution $x$, $s$, $y^+$, $y^-$, $r^+$, and $r^-$, we solve the following linear problem to find the coefficients $b_i^1$, $b_i^2$, and $b_i^3$
\def\arraystretch{1.75}
\begin{equation} \label{eq:inverse}
    \begin{array}{lll}
        \min \quad \phantom{s}& \multicolumn{2}{l}{\mathlarger{\sum}\limits_{i,t}w_t\left((c^1_{it}-\langle Z_{it}, b_i^1 \rangle)^2 + (c^2_{it}-\langle Z_{it}, b^2_i \rangle)^2 + (k_{it}-\langle Z_{it}, b''_i \rangle)^2\right) + \mathlarger{\sum}\limits_{ij} \lambda^j_i ||b_i^j||_1}\\

        \text{s.t.}         & c^1_{it} + 2c^2_{it}x_{it} - p_t^E + \underline \alpha_{it} - \overline \alpha_{it} - \mu_{it} + \mu_{i,t+1} = 0 \phantom{spaace} & \forall i \in \Ic, \forall t \in \Tc \\
                            & p_t^E - \eta \pi_t + \underline \gamma_t^+ - \overline \gamma_t^+ = 0 & \forall t \in \Tc \\
                            & -p_t^E + \eta^{-1} \pi_t + \underline \gamma_t^- - \overline \gamma_t^- = 0 & \forall t \in \Tc\\
                            & \pi_t - \pi_{t+1} + \underline \beta_t - \overline \beta_t = 0 &  \forall t = 1, \dots, T-1 \\
                            & \pi_T + \underline \beta_T - \overline \beta_T = 0 \\
                            & \mu_{it} - \overline \delta_{it} + \underline \delta_{it} + k_{it} = 0 & \forall i \in \Ic, \forall t \in \Tc\\
                            & -\mu_{it} - \overline \theta_{it} + \underline \theta_{it} = 0 & \forall i \in \Ic, \forall t \in \Tc \\
                            & \omega_t = 0, & \forall t \in \Uc(\omega_t), \; \forall \omega \in \{ \underline \gamma^\pm, \overline \gamma^\pm, \overline \beta,  \underline \beta\} \\
                            & \zeta_{it} = 0, & \forall t \in \Uc(\zeta_{it}), \; \forall \zeta \in \{ \underline \alpha, \overline \alpha, \overline \delta,  \underline \delta, \underline \theta, \overline \theta \} \\
                            & \overline \alpha_{it}, \underline \alpha_{it}, \overline \delta_{it},\underline \delta_{it},\overline \theta_{it}, \underline \theta_{it} \leq 0, & \forall i \in \Ic, \forall t \in \Tc \\
                            & \overline \beta_t, \underline \beta_t, \overline \gamma_t^+, \underline \gamma_t^+, \overline \gamma_t^-, \underline \gamma_t^- \leq 0 & \forall t \in \Tc \\
                            & c^2_{it} \geq 0, & \forall i \in \Ic, \forall t \in \Tc.
    \end{array}
\end{equation}
\def\arraystretch{1.}
In the next section, we fit the inverse model \eqref{eq:inverse} to historical price data and then use forward model \eqref{eq:forward_problem} along with the estimated coefficients $b_i$ for prediction.

\section{Results} \label{section:Results}
In this section, we apply the fundamental bottom-up model developed above to the day-ahead electricity markets of Spain, Germany, and France. The primary objective of this section is to evaluate the performance of the proposed inverse optimization approach in forecasting electricity prices out-of-sample. 

We compare our approach to LASSO regression and XGBoost \citep{chen2016xgboost}, two common machine learning methods. We picked LASSO regression as it is a natural simplification of our model. Gradient boosting on the other hand is the state-of-art method for regression problems for medium-sized tabular datasets and is widely used in practice. To compare the methods, we report the results of extensive out-of-sample experiments. All models have been implemented in Python using Gurobi 9.5 for the inverse and forward problems and \emph{scikit-learn} \citep{pedregosa2011scikit} for LASSO regression and XGBoost.

\subsection{Data \& Features}
We obtain data on spot prices, installed capacity, renewable generation, and demand from January 2015 to February 2022 from the \emph{European Association for the Cooperation of Transmission System Operators for Electricity} (ENTSO-E). For all models presented in this paper, the demand data is adjusted by subtracting both cross-border transactions and all forms of renewable generation from the original total demand, i.e., we are working with \emph{residual demand}. A summary of the ENTSO-E data for the evaluation periods is given in Table \ref{table:countries_charac}, with all metrics indicating average values across each period. 
The data sourced from ENTSO-E encompasses a variety of energy technologies in Spain, Germany, and France. Common technologies across these markets include power plants that operate on biomass, lignite, coal, gas, oil, nuclear, solar, onshore wind, and hydropower, which encompasses pumped storage, run-of-river, and water reservoir systems. Additionally, the German system has power plants with offshore wind and geothermal capacities. For the purposes of our analysis, we consider marginal costs, ramping costs, and ramping restrictions only for conventional, non-renewable power plants, namely lignite, coal, gas, oil, and nuclear facilities.

\begin{table}[t]
    \centering
    \setlength{\extrarowheight}{7pt}
    \small 
    \begin{tabular}{p{2.5cm}clcccc} 
        \toprule
        \multicolumn{1}{l}{Periods} & Country & \multicolumn{1}{p{1.5cm}}{\centering Demand (MW)} & \multicolumn{1}{p{1.1cm}}{\centering Price (\euro)} & \multicolumn{1}{p{2cm}}{\centering Renewable Penetration\\(\%)} & \multicolumn{1}{p{1.7cm}}{\centering Import (MW)} & \multicolumn{1}{p{1.7cm}}{\centering Export (MW)} \\
        \midrule
        \multirow{3}{*}{Jan 18 - Dec 20} & Spain & 27119 & 46.3 & 0.43 & 1046 & 1972 \\
        & Germany & 55597 & 37.5 & 0.48 & 7013 & 2906 \\
        & France & 52316 & 40.6 & 0.25 & 6098 & 1211 \\
        \midrule
        \multirow{3}{*}{Jan 19 - Dec 21} & Spain & 26909 & 64.5 & 0.46 & 1354 & 1755 \\
        & Germany & 54818 & 55.0 & 0.49 & 6352 & 3276 \\
        & France & 51952 & 60.2 & 0.25 & 5833 & 1621 \\
        \midrule
        \multirow{3}{*}{Apr 21 - Jan 22} & Spain & 26767 & 140.6 & 0.48 & 1668 & 1581 \\
        & Germany & 54786 & 117.9 & 0.47 & 5923 & 3134 \\
        & France & 50506 & 136.2 & 0.24 & 5374 & 2058 \\
        \bottomrule
    \end{tabular}
    \caption{\label{table:countries_charac} Characteristics of studied markets in the three out-of-sample periods. The table reports averages over the hours in the respective periods. Renewable penetration refers to the share of domestic demand (without taking into account imports and exports) covered by renewable generation.}
\end{table}

Additionally, we consider a range of features $Z = Z_1 = Z_2 = Z_3$ to forecast the cost structures \eqref{eq:regression} in our inverse optimization model \eqref{eq:inverse}. In particular, we use hourly demand, fuel and emission prices, temperature, renewable generation from wind, solar, and run-of-the-river plants, as well as holidays as features. We create dummy variables for hour of the day, weekdays, and holidays. Moreover, we generate second order interactions by multiplying all features with one another. As our machine learning models all use regularization, we use min-max scaling to normalize features to unit scale. Spot prices for fossil fuels and emission allowances as well as hourly temperature for each country are obtained through \emph{Refinitiv EIKON}. Table~\ref{table:Market_Features} presents a comprehensive overview of features we utilized in all three regression models as described in \eqref{eq:regression} and for all countries.

\begin{table}[t]
    \centering
    \begin{tabular}{l l r r r r r r}
        \toprule
        Feature & Unit & \multicolumn{2}{c}{Spain} & \multicolumn{2}{c}{Germany} & \multicolumn{2}{c}{France} \\
        \cmidrule(r){3-4} \cmidrule(r){5-6} \cmidrule(r){7-8}
               &      & \multicolumn{1}{r}{Mean} & \multicolumn{1}{r}{Std} & \multicolumn{1}{r}{Mean} & \multicolumn{1}{r}{Std} & \multicolumn{1}{r}{Mean} & \multicolumn{1}{r}{Std} \\
        \midrule
        Actual Total Load & MW & 27880 & 5760 & 57256 & 10036 & 53617 & 11828 \\
        Natural Gas Price (TTF) & EUR/MWh & 23 & 22 & 23 & 22 & 23 & 22 \\
        Thermal Coal Price (API2) & EUR/t & 78 & 39 & 78 & 39 & 78 & 39 \\
        Carbon Price & EUR/t & 22 & 20 & 22 & 20 & 22 & 20 \\
        Temperature & \textdegree C & 16 & 6 & 10 & 7 & 13 & 6 \\
        Solar Generation & MW & 1771 & 2327 & 4541 & 7049 & 1160 & 1659 \\
        Wind Onshore Generation  & MW & 5730 & 3499 & 10075 & 8362 & 3235 & 2524 \\
        Run-of-river Generation & MW & 985 & 406 & 1644 & 386 & 4584 & 1497 \\
        Day of the Week & Categorical & - & - & - & - & - & - \\
        Hours of Day & Binary & - & - & - & - & - & - \\
        Holidays & Binary & - & - & - & - & - & - \\
        \bottomrule
    \end{tabular}
    \caption{\label{table:Market_Features} Overview of descriptive statistics for features used in \eqref{eq:regression}.}
\end{table}

As the inverse optimization model does not estimate ramping limits but only ramping cost, we set these limits for every technology $i$ to the largest observed ramps between successive periods for this technology for time $t$ in the training period: $R^+_{i} = \max_{t} \max(x_{it}-x_{i,t-1},0)$,  $R^-_{i} = \max_{t} \max(x_{it}-x_{i,t-1},0)$, and $R^+_i = \max_{t} R^+_{it}$. To parameterize storage, we obtain energy capacities $\bar S_t$ from ENTSO-E and find the power capacities as the maximum observed power input and output for times $t'$ in the training period as $\bar Y_t^+ = \max_{t'} \; y_{t'}^+$ and $\bar Y_t^- = \max_{t'} \; y_{t'}^-$.

We partition our dataset to define three experiments $E_1$, $E_2$, and $E_3$. For the first experiment $E_1$, we use the three years between Jan 2015 and Dec 2017 as training data and the three years between Jan 2018 and Dec 2020 as the test data. Similarly, we define the training and test periods for experiment $E_2$ as Jan 2016 to Dec 2018 and Jan 2019 to Dec 2021. Finally, for $E_3$ the data between Jan 2015 and Mar 2021 is the training data and Apr 2021 to Jan 2022 is the test data. The relatively long three years evaluation period of the first experiment contains two pre-COVID years and the first year of the COVID pandemic featuring plummeting prices for all energy commodities accompanying the decrease in global industrial output. The second experiment features two COVID years in the test data, where the second year is characterized by the gradual normalization of energy markets. Finally, the third period includes the surge in fuel prices in 2021 due to the start of the Russo-Ukrainian war in the test data. In summary, the three experiments contain a range of distribution shifts in energy markets and therefore represent an ideal testbed for long-term forecasting methods that have to be able to forecast prices in regimes that are structurally different from the ones observed in the training data.

\subsection{Performance Metric}
To compare the accuracy of our predictions with the benchmarks, we use a normalized version of the mean absolute error (NMAE), which we define as follows,

$$\text{NMAE}:= \frac{\sum_{t=1} w_t |\hat p^E_t -p^E_t|}{\bar p^E {\sum_{t=1}^T w_t}}.$$
Here, $p^E_t$ is the electricity price in period $t$, $\hat p^E_t$ is its estimate by the inverse model or the employed machine learning methods, $w_t$ is the weight, and $\bar p^E= T^{-1} \sum_{1}^T p^E_t$ is the average price. If we set the weight to the production of the renewable plant in period $t$, i.e. $w_t = q_t$, we obtain the NMAE for the capture price. If we set the weight equal to one, i.e., $w_t \equiv 1$, we obtain the NMAE for the base price. 

The choice of using the NMAE as a performance metric is motivated by its ability to give higher weights to errors in periods when production is high and lower weights when there less production. This information is pertinent to pricing PPAs for solar projects, which capture the highest value around noon and no value at night. Additionally, by normalizing the absolute error with respect to the average price, the NMAE allows for more meaningful comparisons across different test sets with different average price levels.

\subsection{Hyperparameter Tuning}
All the employed models have hyperparamters that cannot be directly estimated along with the other model parameters. For every of our experiments $E_1$, $E_2$, and $E_3$ and every method we separately tune the hyperparameters using 5-fold cross validation with $20\%$ of the data in the validation data set and the weighted MSE with weights $w_t$ as error measure. Correspondingly, the chosen loss function for the LASSO and XGBoost are also the weighted MSE.

In particular, for the inverse optimization models the regularization parameters $\lambda_i^j$ have to specified before the model is estimated. Similarly, for the LASSO regression, the regularization parameter has to be chosen, and XGBoost has a range of hyperparameters. For the latter we tune the three parameters \texttt{eta, maxdepth, alpha} and let the \emph{scikit-learn} implementation of XGBoost choose the other parameters automatically. 

To find suitable parameters for LASSO and XGBoost, we perform a grid search over the parameter space. For inverse optimization, we must select a regularization parameter for marginal as well as ramping costs for each technology, which amounts to choosing 15 parameters for $\lambda^1$, $\lambda^2$, and $\lambda^3$. As doing a grid search is prohibitive in this case, we use \cite{hansen2001cma}'s covariance matrix adaptation evolution strategy (CMA-ES) to search for the best regularization parameters. 

\subsection{Out-of-Sample Results}
Performance metrics for the evaluation periods are shown in Table~\ref{table:OoS_Long_Term} grouped by NMAEs for the average price (\emph{base}), solar production (\emph{solar}), and wind production (\emph{wind}). The difference in between these experiments is how the weights $w_t$ are chosen: For \emph{base} we choose $w_t=1$ for all $t \in \Tc$, while for \emph{solar} and \emph{wind} we choose the corresponding capacity factors, i.e., the actual production divided by the installed power capacity, of the technology in question in the corresponding country. The lowest NMAE within each group is highlighted in boldface.

\begin{table}[]
    \centering
    \setlength{\extrarowheight}{7pt}
    \begin{tabular}{llccccccccc}
        \toprule
        &         & \multicolumn{9}{c}{Normalized MAE} \\
        &         & \multicolumn{3}{c}{Spain} & \multicolumn{3}{c}{Germany} & \multicolumn{3}{c}{France} \\
        \cmidrule(lr){3-5} \cmidrule(lr){6-8} \cmidrule(lr){9-11}
        Experiment                                & Method   & Base    & Solar    & Wind    & Base    & Solar    & Wind     & Base    & Solar    & Wind    \\ 
        \midrule
        \multirow{3}{*}{$E_1$}                   & InvOpt     & \textbf{0.12}    & \textbf{0.10}   & \textbf{0.12}   & \textbf{0.17}    & \textbf{0.15}    & \textbf{0.15}    & \textbf{0.20}    & \textbf{0.18}    & \textbf{0.19}   \\
                                                       & LASSO     & 0.50    &0.37    &0.53    & 0.25    & 0.24    & 0.24    & 0.21    & 0.20    & \textbf{0.19}   \\
                                                       & XGBoost &  0.18    &0.12    &0.13    &0.29     &0.22    &0.22     & 0.32   & 0.23    & 0.30   \\ 
        \midrule
        \multirow{3}{*}{$E_2$}                   & InvOpt     & \textbf{0.20}    & \textbf{0.17}   & \textbf{0.18}   & \textbf{0.18}    & \textbf{0.16}    & \textbf{0.18}    & \textbf{0.24}    & \textbf{0.20}    & \textbf{0.22}  \\
                                                       & LASSO     & 0.29    &0.27    & 0.28   & 0.26    &  0.29   & 0.34    & 0.29    &  0.27   & 0.28   \\
                                                       & XGBoost & 0.41    & 0.34   & 0.34   &0.43     &0.37     &0.34    &0.46     &0.36     &0.34    \\ 
        \midrule
        \multirow{3}{*}{$E_3$}                  & InvOpt     & \textbf{0.18}    & \textbf{0.16}   & \textbf{0.16}   & \textbf{0.22}    & \textbf{0.18 }   & \textbf{0.22}    & \textbf{0.19}    & \textbf{0.15}    & \textbf{0.17}   \\
                                                       & LASSO     & 0.32    & 0.27   & 0.35   & 0.33    &  0.33   &  0.32   & 0.59    & 0.57    & 0.54   \\
                                                       & XGBoost & 0.58    & 0.50   & 0.47   & 0.62    & 0.58    & 0.55    & 0.61    & 0.53    & 0.54  \\
        \bottomrule
    \end{tabular}
    \caption{\label{table:OoS_Long_Term} NMAE for the electricity base price as well as wind and solar capture prices for the three experiments $E_1$, $E_2$, and $E_3$ in Spain, Germany, and France.}
\end{table}

As we can see, inverse optimization (InvOpt) generally outperforms the other two machine learning methods by quite some margin. We also find that althoug the inverse optimization model beats the machine learning methods across the board, it performs relatively better in forecasting the capture prices, in particular solar, than forecasting base electricity prices, which makes it a good model for PPA pricing. When looking at the three different countries, it seems that the inverse optimization model performs best for the German market and worst in France (with the exception of $E_3$). 
When looking at the three different countries, it seems that the inverse optimization model performs best for the Spanish market, with the exception of $E_2$, and worst in France, with the exception of $E_3$.
Regarding the comparison of the two machine learning benchmarks, we note that, surprisingly, XGBoost performs better than LASSO only for $E_1$ in Spain and is otherwise significantly outperformed by the much simpler LASSO regression. 

Unlike the two machine learning benchmarks, the inverse optimization model captures relevant physical characteristics as well as the market-clearing mechanism of merit-order-based electricity markets. The structural information included in the forward model therefore helps dealing with distribution shifts, i.e., situations when training data is sampled from a different distribution than test data. 

\begin{figure}[t]
    \centering
    \includegraphics[width=1\textwidth]{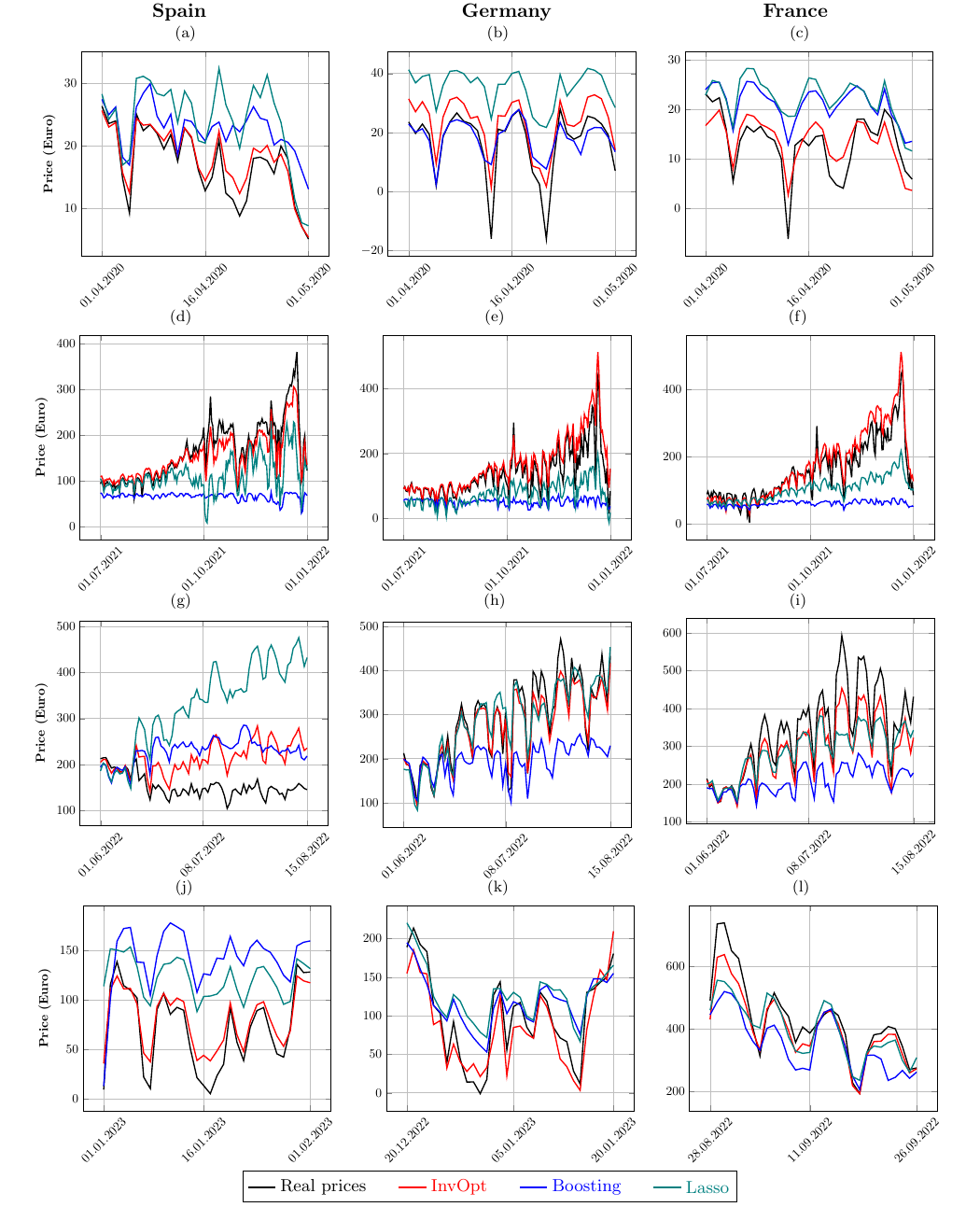}
    \caption{Daily Average of Day-Ahead Electricity Prices over Selected Periods. 
    }
    \label{fig:events_ES}
\end{figure}

Next, we investigate more closely the time periods where such distribution shifts occurred. For example, this was the case in early 2020 when extensive lockdowns led to a plunge in electricity demand across the world. In 2021, there were notable spikes in fuel prices. This was followed by another significant shift in the middle of 2022 due to the onset of the Russo-Ukraine war, which sent natural gas prices skyrocketing. Other situational shifts, like high wind generation in Germany, increased hydro capacity in Spain, and a decline in nuclear output in France, also significantly influenced electricity prices in their respective zones. Figure \ref{fig:events_ES} presents the predicted and actual prices, focusing on the daily averages of hourly day-ahead electricity prices in these periods, displaying Spanish prices in the first column, German prices in the second, and French prices in the third.

The first row of Figure \ref{fig:events_ES}, consisting of panels (a), (b), and (c), illustrates the day-ahead electricity prices and their predictions for April 2020, the first month of the COVID-19 lockdowns in Europe. This period was the most dramatic in terms of consumption and price changes on the energy markets. Except for Germany, the machine learning benchmarks tend to show a marked upward bias during this time period and are not able to handle the change in circumstances well. The inverse optimization model, by contrast, while being also slightly upward biased, by and large predicts prices correctly.

Regarding the 2021 gas price hike and its impact on electricity prices, the second row of Figure \ref{fig:events_ES} (panels d, e, and f) shows a striking difference between the machine learning methods and inverse optimization, with LASSO and XGBoost severely underestimating prices and thus not being able to predict the massive price hikes caused by the gas shortage at that time. In contrast, the inverse optimization model deals with this situation quite well and predicts high prices that are close to actually realized market prices.

The impact of the start of the Russo-Ukrainian war is examined in the third row of Figure \ref{fig:events_ES} (panels g, h, and i). This section reveals that in Spain, neither method could predict prices accurately due to the country's capping of gas prices. However, both inverse optimization and to a lesser degree LASSO regression are effective in tracking prices in Germany and France, unlike gradient boosting, which falls short.

Panel (j) in Figure~\ref{fig:events_ES} explores a period of one month in Spain characterized by high hydropower production. Again, only the inverse optimization model is able to deal with the unusual situation and predict prices well.

For Germany, Panel (k) of Figure~\ref{fig:events_ES} assesses a period of one month with increased wind power production. While the price effect is not as pronounced as in Spain's hydropower scenario, inverse optimization remains closely aligned with the actual prices, unlike LASSO and XGBoost, which both underestimate the price-dampening effect of high wind power output.

Lastly, Panel (l) of Figure~\ref{fig:events_ES} focuses on a period of low nuclear power output in France. In this instance, while the differences between the models are not as marked as in other periods, XGBoost exhibits a tendency towards downward-biased price predictions, while the predictions of LASSO exhibit large errors in both directions with the inverse optimization model again producing the most reliable forecasts.

In light of the above, we conclude that inverse optimization is better able to handle distributional shifts than standard machine learning methods, since the fundamental bottom-up model enables it to  account for the non-linearity of the merit order as well as the effect of exogenous factors on the marginal cost of each technology. Clearly, this is important when studying long-term PPA contracts that extend far into the future when the electricity sector as a whole will most likely be significantly different from the present situation or anything that can be directly forecast from historic data without a mediating system model.

\section{Case Study} \label{section:CaseStudy}
In this section, we present a hypothetical case study of a steel company that seeks to supply one of its Spanish production sites with renewable power. The case is inspired by our discussions with \emph{ArcelorMittal} on similar contracts. We use this case study to illustrate how the model can be used as a source of market intelligence, supporting the negotiation process with the developer and enabling the buyers to make a more informed decisions.

We assume that the buying company evaluates a PPA with the developer of a solar power plant in the same region. The hypothetical solar power plant consists of multiple arrays of photovoltaic panels with an overall installed power capacity of $50$ MW. The offered PPA encompasses the entire power production for a fixed term of 7 years. We suppose the buyer reviews the PPA in February 2022.

After having reviewed the specifications of the PPA, the buyer uses the proposed inverse optimization model to calculate the expected capture price of the PPA. The calculation comprises the following steps:
\begin{enumerate}
    \item After acquiring the necessary historical data and features, the buyer fits the inverse optimization model outlined in Section~\ref{ssec:inverse_opt} to historical data, in this case spanning a time period from January 1, 2015, to February 1, 2022.
    
    \item Based on market data, reports, and internal models, the buyer then generates one or more scenarios for the possible evolution of exogenous factors that enter the forward model. The required information concerns future installed capacities for the different technologies, fuel and carbon prices, as well as demand.
    
    Market data and reports can be obtained from governmental infrastructure plans, data service providers such as \emph{ICIS}, \emph{Refinitiv}, or \emph{Platts}, as well as energy news and reports from \emph{Montel} or other industry-specific publications.
    
    \item Using a scenario about the development of capacities, prices, demand, and renewable generation, the buyer uses estimated parameters from step 2 to forecast marginal cost and ramping cost over the seven-year term of the PPA. 
    
    \item Finally, the buyer uses the forward model to obtain an estimate for the day-ahead prices for each scenario. These prices can then be used to calculate the forecast capture price of the project for the given scenario according to \eqref{eq:capture_price}.
\end{enumerate}

In the next section, we detail this workflow based on the concrete project outlined above and real-world data on the electricity sector in Spain.

\subsection{Input Data \& Future Scenarios}
To create a scenario of the development of installed capacity and demand, we use data from Spain’s National Energy and Climate Plan (NECP) from January 2020. The NECP's projections for 2030 anticipate a total installed power capacity of 161 GW, which will comprise 50 GW wind power, 46 GW solar power, 27 GW combined cycle gas, 15 GW hydro, 9.5 GW pumped-hydro, 7 GW solarthermal, and 3 GW nuclear. Note that in particular, Spain's NECP forecasts a complete phase-out of coal in the power sector by 2030. Additionally, the plan includes some marginal technologies including biomass, oil, other renewable energy sources not explicitly listed, energy from waste, and combined heat and power~\citep{ec2020}. However, for the purposes of this case study, these technologies are not considered due to their minor contribution to overall capacity. We also do not consider cross-border capacities between Spain and Portugal and Spain and France, as the focus is primarily on Spain's domestic energy capacity and demand. 

We assume a linear growth/de-growth trajectory from present capacities to the anticipated levels in 2030.  In line with this approach, the calculated ramping capacities for conventional, non-renewable power plants such as gas, coal, and nuclear are adjusted to grow or decline at the same percentage rate as their installed capacities. Further details are provided in Table~\ref{table:demandcapacities}.

\begin{table}[t]
    \centering
    \begin{tabular}{cccccccccc}
        \toprule
        \multirow{2}{*}{Year} & \multirow{2}{*}{Demand (TWh)} & \multicolumn{7}{c}{Installed Generation Capacity (GW) by Technology} \\
        \cmidrule(lr){3-9}
        & & Solar & Wind & Hydropower & Gas & Coal & Nuclear & Pumped-storage \\
        \midrule
        2023 & 247.64 & 24.00 & 30.00 & 16.00 & 29.90 & 3.22 & 7.10 & 3.42 \\
        2024 & 255.07 & 27.14 & 32.86 & 16.00 & 29.49 & 2.76 & 6.51 & 4.29 \\
        2025 & 262.72 & 30.29 & 35.71 & 16.00 & 29.07 & 2.30 & 5.93 & 5.16 \\
        2026 & 270.60 & 33.43 & 38.57 & 16.00 & 28.66 & 1.84 & 5.34 & 6.03 \\
        2027 & 278.72 & 36.57 & 41.43 & 16.00 & 28.24 & 1.38 & 4.76 & 6.89 \\
        2028 & 287.08 & 39.71 & 44.29 & 16.00 & 27.83 & 0.92 & 4.17 & 7.76 \\
        2029 & 295.70 & 42.86 & 47.14 & 16.00 & 27.41 & 0.46 & 3.59 & 8.63 \\
        2030 & 304.57 & 46.00 & 50.00 & 16.00 & 27.00 & 0.00 & 3.00 & 9.50 \\
        \bottomrule
    \end{tabular}
    \caption{Yearly Electricity Demand and Power Generation Capacity in Spain from 2023 to 2030. The column ``Demand (TWh)'' shows the yearly 3\% growth of electricity demand from the current quantity. 2030 values are based on the NECP from 2020, 2023 values are taken from ENTSO-E database and the capacities in intermediary year are calculated by interpolation.}
    \label{table:demandcapacities}
\end{table}

\begin{figure}[ht]
    \centering
    \includegraphics[width=1\textwidth]{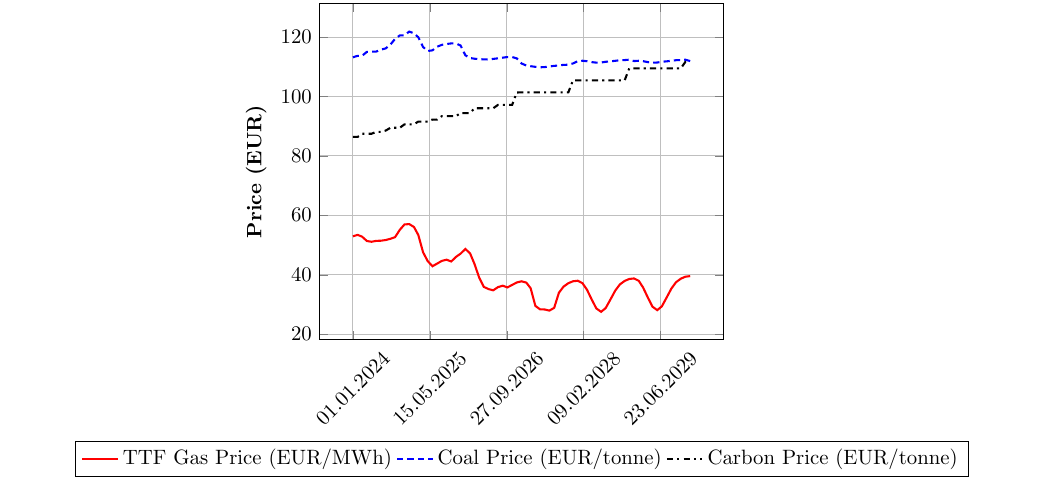}
    \caption{Long-Term Forecast of Gas, Coal, and Carbon Prices (2023-2030) based on Price Forward Curves from Refinitiv EIKON.}
    \label{fig:forward_curve_fuels}
\end{figure}

Long-term price forecasts for fuel and carbon prices from 2023 to 2030 are sourced based on price forward curves provided by Refinitiv EIKON. Figure \ref{fig:forward_curve_fuels} shows the forecasts of the relevant price factors.

Based on this data, we create three scenarios that outline different possible future developments of the electricity sector in Spain: an ambitious scenario where Spain fulfills all its current climate related goals, a business-as-usual scenario where the implementation of these goals is sluggish, as well as an intermediate scenario which represents a middle ground between the two extremes. These scenarios are merely intended to exemplify how a what-if analysis about future developments can be used in PPA pricing and should not serve as realistic predictions, which would require a much more in-depth analysis of the Spanish power sector and is beyond the scope of this paper.

\begin{enumerate}
    \item \textbf{Ambitious Scenario}
    \begin{itemize}
        \item[-] The NECP will be fully implemented, resulting in a total installed power capacity as foreseen by the plan for the year 2030. Table~\ref{table:demandcapacities} presents the yearly capacities under this scenario.
        
        \item[-] Fuel prices and emission allowances are projected to undergo a gradual percentage change, linearly interpolating from 0\% to an anticipated decline of 10\% by the final year of 2030, relative to the monthly price forward curves in Fig.~\ref{fig:forward_curve_fuels}. This trend is attributed to a surge in renewable energy adoption. 
        
        \item[-] We assume a rapid electrification of heating and transport. Correspondingly, we consider a yearly 3\% growth of electricity demand until 2030~\citep{McKinsey2024EuropeanPower}. Table~\ref{table:demandcapacities} presents the yearly demand under this scenario.
    \end{itemize}

    \item \textbf{Business-as-Usual Scenario}
    \begin{itemize}
        \item[-] We assume that for renewable power plants, only 30\% of the additional renewable capacity planned in the NECP will be achieved by 2030. For non-renewable power plants, such as gas and nuclear, installed capacities have been calculated to maintain a constant ratio of yearly demand to firm production capacities, as seen in the ambitious scenario. In particular, to compensate for the capacities of the renewables that are not developed, we phase out less conventional production to reach the same ratio of demand to firm capacities.\footnote{Based on the installed capacities and actual generation of power plants in Spain in 2022, we have calculated the firm capacity factors for various technologies as follows: Solar at 25.5\%, Wind at 23.5\%, Hydropower at 12.1\%, Gas at 31.0\%, Coal at 19.2\%, Nuclear at 90.0\%, and Hydro pumped-storage at 24.4\%.} To this end, we prioritize nuclear power plants and then gas reach the decided firm capacities in 2030. It turns out that these two technologies are sufficient and it is possible to fully phase out coal also in the BAU scenario. We again use linear interpolation to calculate the capacities for the years between 2024-2029. 
        
        \item[-] For this scenario, our projections for monthly fuel prices and emission allowances from 2023 to 2030 align precisely with the data obtained from Refinitiv EIKON (Fig.~\ref{fig:forward_curve_fuels}).
        
        \item[-] Since electrification is sluggish in this scenario, we only assume a modest annual average growth rate of 1\% for yearly electricity demand from 2022 to 2030.
    \end{itemize}

    \item \textbf{Intermediate Scenario}
    \begin{itemize}
        \item[-] This scenario assumes that by 2030, 60\% of the additional renewable capacity planned in the NECP will be achieved. Similar to the Business-as-Usual scenario, for non-renewable sources such as gas and nuclear, capacities have been calculated by maintaining a constant ratio of demand to firm capacities and utilizing capacity factors for production estimations.

        \item[-] We anticipate a gradual percentage shift in monthly fuel and emission prices, with a linear decrease from the current levels, reaching a minor escalation of 5\% by the final year of 2030, compared to the monthly price forward curves in Fig.~\ref{fig:forward_curve_fuels}. 

        \item[-] For this scenario, we project a 2\% yearly average increase in electricity demand throughout until 2030.
    \end{itemize}
\end{enumerate}

\subsection{Models for Wind and Solar Capacity Factors}
To derive hourly data for production from solar, wind, and run-of-river generation, as well as temperature and demand, we use a series of regression models that capture yearly, weekly, and daily seasonality.

Denote hours \(\mathcal{H}\) as the set of days that are holidays, \(\mathcal{W}_j\)  as the set of days that are weekday $j$, $h$ as the hour of the day and $d$ as a day from the training and scenario data. 

For power generation from solar, wind, and run-of-river (as well as temperature) the regression model only considers seasonal variation using multiple trigonometric terms with a maximum cycle length of 365 days. We fit a separate model for each hour of the day. The resulting regression model is given by
\[
Y_{dh} = \beta^0_{h} + \beta^1_{h}d + \sum_{i = 1}^{180}  \left( \beta^4_{hi} \sin\left(\frac{di 2\pi}{365}\right) + \beta^5_{hi} \cos\left(\frac{di 2\pi}{365}\right)\right).
\]

The demand model additionally includes calendar features using dummy variables for holidays and day of the week. The resulting regression model is given by
\[
D_{dh} = \beta^0_{h} + \beta^1_hd + \sum_{j=1}^6 \beta^2_{hj} \textbf{1}_{\mathcal{W}_j}(d)   + \beta^3_{h} \textbf{1}_{\mathcal{H}}(d) + \sum_{i = 1}^{180}  \left( \beta^4_{hi} \sin\left(\frac{di 2\pi}{365}\right) + \beta^5_{hi} \cos\left(\frac{di 2\pi}{365}\right)\right).
\]

All features are pre-processed using quantile transformation and expanded to second-order polynomials to obtain interaction terms. Models are then fitted using cross-validated Lasso regression. 

In Table~\ref{table:model_fits}, we present the fits for each regression model, including the Root Mean Square Error (RMSE), Mean Absolute Error (MAE), and the R$^2$, along with the selected regularization parameter for LASSO regression. In our analysis, we use scikit-learn version 1.1.3 and allow LASSO to optimize the regularization parameter using 5-fold cross-validation with $20\%$ of the data in the corresponding validation sets.

\begin{table}[t]
    \centering
    \begin{tabular}{lccccc}
        \toprule
        \textbf{Model} & \textbf{Selected $\lambda$ for LASSO} & \textbf{RMSE} & \textbf{MAE} & \textbf{R\textsuperscript{2}} \\
        \midrule
        Solar Capacity Factor & $6.61596 \times 10^{-4}$
         &0.72 &0.39 &0.61 \\
        Wind Capacity Factor &$6.30053 \times 10^{-2}
        $ &1.13 &0.82 &0.02 \\
        Run of River Capacity Factor &$6.69094 \times 10^{-3}
        $ &0.97 &0.61 &0.29 \\
        Temperature &$5.87843 \times 10^{-4}
        $ &0.42 &0.32 &0.82 \\
        Demand &$1.48617 \times 10^{-4}
        $ & 0.86 &0.49 &0.44 \\
        \bottomrule
    \end{tabular}
    \caption{\label{table:model_fits} Model Fit Metrics and Selected Lambda for LASSO Regression}
\end{table}

Inspecting the results in Table \ref{table:model_fits}, we observe that the models for solar, temperature, and demand exhibit decent fits, while the models for run-of-river and wind capacity factors fit rather poorly, which indicates that they cannot be explained well by seasonal factors.

\subsection{Results}
By combining solar generation forecasts with forecast prices in prices \eqref{eq:capture_price}, we can calculate the daily average capture prices for each scenario. Additionally, we calculate the net present value of the PPA by discounting the cash flows resulting from generating power at the given hourly prices minus the fixed price that will be paid to the developer. We use an annual discount rate of $11\%$ to reflect the cost of capital of the company. 

We determine the break-even PPA prices for each scenario, which are the PPA prices at which the net present value of the PPA equals zero. More specifically, we calculated the break-even PPA prices, $P_{PPA}$, by solving the equation 
$$\sum_{t\in \Tc} \frac{Q_t \times (P_{PPA}-\hat p_t^E)}{\rho_t} = 0,$$
where $Q_t$ is the hourly electricity production. The resulting PPA prices are \euro{125.93}, \euro{100.81}, and \euro{116.04} for the BAU, ambitious, and intermediate scenario, respectively. 

\begin{figure}[ht]
    \centering
    \includegraphics[width=1\textwidth]{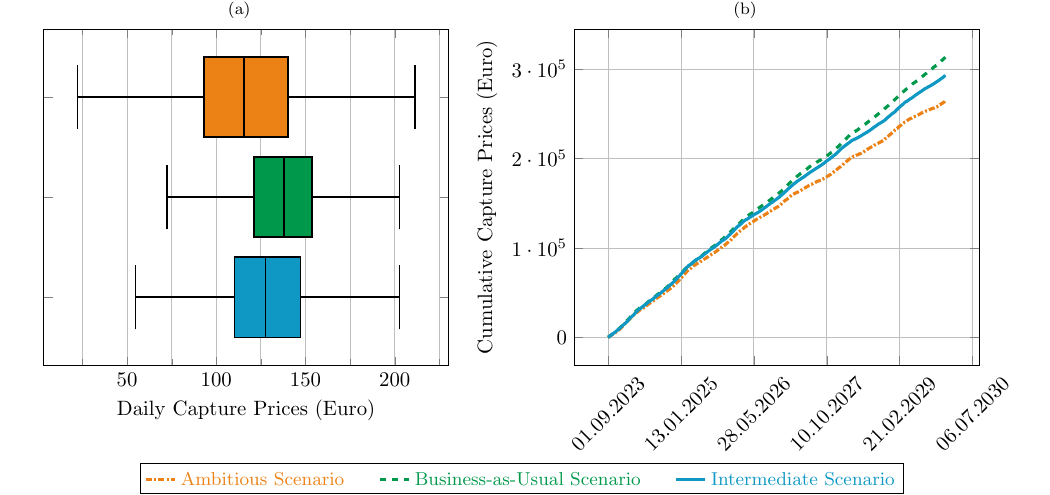} 
    \caption{Analysis of solar capture prices from 2023 to 2030: (a) Distribution of daily solar capture prices, and (b) Cumulative solar capture prices under different scenarios}
    \label{fig:caseStidy_results}
\end{figure}

The analysis of solar capture prices from 2023 to 2030, including the distribution of daily solar capture prices and the cumulative solar capture prices under different scenarios, is presented in Figure~\ref{fig:caseStidy_results}.

\subsection{Sensitivity Analysis}
The above analysis offers an estimate of the capture price for a fixed set of scenarios. However, since the scenarios depends on many uncertain parameters, it is also important to understand the sensitivity of the capture price to changes in input factors. We therefore additionally conduct a sensitivity analysis by changing the values of relevant input factors ceteris paribus to study their relative effect on the capture price.

\begin{figure}[t]
    \centering
    \includegraphics[width=1\textwidth]{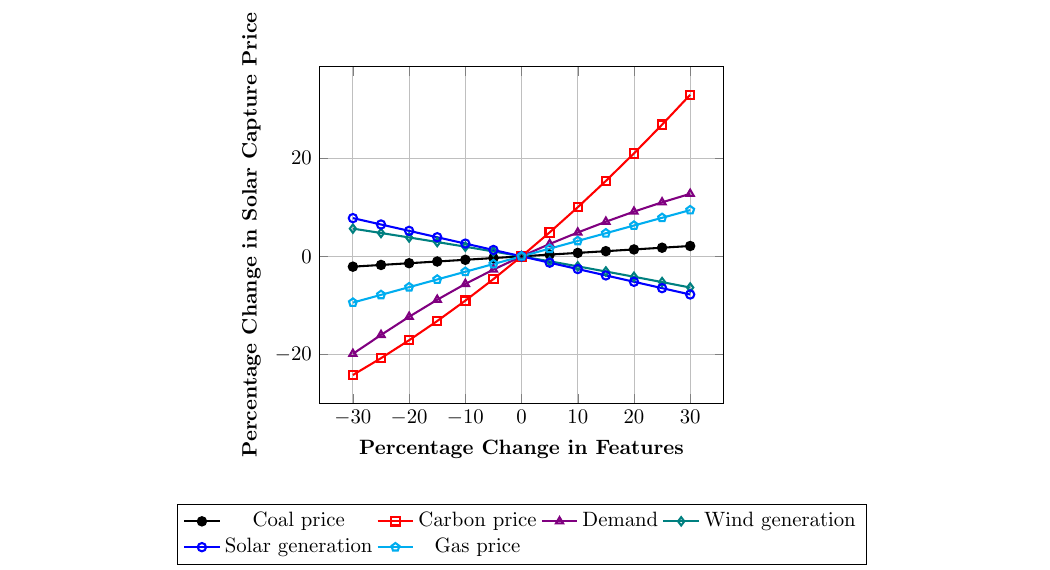} 
    \caption{\label{fig:sensitivity_analysis} Sensitivity of the capture price with respect to exogenous input factors.}
\end{figure}

The business-as-usual scenario serves as the base case for this analysis. We modify factors, such as gas price, coal price, carbon price, demand, and the outputs of wind and solar generation (leaving the output of the solar plant contracted in the PPA constant). For each of these factors, we create 13 equi-distant points, spanning an interval that ranges from 30\% below to 30\% above the base case values. We then compare the variation in the capture price against the capture price established in the business-as-usual scenario by measuring the percentage change in the corresponding input parameters. The result of the sensitivity analysis is shown in Figure~\ref{fig:sensitivity_analysis}. 

As we can see, carbon price and demand have the highest positive correlation on the capture price, followed by gas price, while solar power output exhibits the highest negative effect. Both of these make intuitive sense: Increases in solar output lead to cannibalization effects that render the PPA less profitable, whereas changes in fuel and emission prices effect the market clearing price that is defined by the most expensive power plant, which during peak hours where solar output is highest, is often a gas-fired power plant. Changes in coal price rate have the smallest effect, which is predictable given the coal phase-out by 2030.

The sensitivity analysis above allows the interpretation of the different capture prices for the three scenarios discussed in the last section. For example, in the ambitious scenario that features the lowest capture price, the combined negative effects of higher solar and wind generation capacity, coupled with lower fuel and emission prices, outweigh the impacts of increased demand, leading to the scenario's lowest cumulative prices. This shows that using more renewable energy and less expensive fuels can help keep prices low, even when demand goes up. 
Moreover, as we see in Figure~\ref{fig:caseStidy_results}(a), the box plot relevant to the ambitious scenario has the highest interquartile range, which indicates more variable prices due to higher renewable penetration in the grid.

Conversely, the business-as-usual scenario exhibits the highest cumulative prices, which can be attributed to the combination of the highest fuel and emission prices along with the lowest installed capacities for renewable energies. These factors have a stronger upward influence on cumulative prices than the impact of having the lowest demand when compared to the other two scenarios. Additionally, the narrowest interquartile range observed in the box plot for the business-as-usual scenario in Figure~\ref{fig:caseStidy_results}(a) corresponds to the lowest renewable power outputs relative to the other two scenarios.
\section{Conclusion} \label{section:Conclusion}
We introduce a novel approach to valuing Power Purchase Agreements (PPAs) for corporate renewable energy procurement, blending fundamental electricity market models with statistical learning techniques. Our model, which utilizes regularized inverse optimization of a quadratic programming formulation of a fundamental bottom-up model, represents an innovation over previous approaches in predicting electricity market prices and, by extension, the valuation of PPAs.

We provide extensive evidence, covering market data from Spain, Germany, and France that demonstrate the superior performance of our approach compared to traditional statistical learning benchmarks. Particularly notable is the model's ability to handle distribution shifts in the market, such as those caused by the COVID-19 pandemic and the 2021 gas price hike. The ability to handle distribution shifts is important for predicting the price effect resulting from market fundamentals, such as fuel cost and generation mix.

In a detailed case study, we illustrate the practical application of our model in a real-world setting. By generating custom scenarios and conducting sensitivity analyses, we demonstrate the model's capability to provide valuable insights for buyers of PPAs. This is especially useful for companies that seek to align their energy procurement strategy with corporate sustainability goals.

Future research could extend this model's capabilities by incorporating stochastic elements to better capture the uncertainties in market drivers and exploring its applicability in a broader, pan-European context. This would further enhance the model's scope and utility for market players dealing with PPAs.

\ACKNOWLEDGMENT{This publication immensely benefited from the close cooperation with the analytics team of ArcelorMittal Energy SCA in particular Tom Bausch, Ashank Sinha, and Can Turktas who helped the authors understand the intricacies of the European spot markets and provided the data to run the analyses.  This research is supported by the National Research Fund of Luxembourg through grant 13771994.}

\bibliographystyle{pomsref} 
\bibliography{arXive_submission} 
\end{document}